\def\aa{{A\&A}}
\def\aas{{A\&AS}}
\def\aj{{AJ}}
\def\annrev{{ARA\&A}}
\def\apj{{ApJ}}
\def\apjs{{ApJS}}

\def\mnras{{MNRAS}}
\def\nat{{Nature}}
\def\pasp{{PASP}}
\newcommand{\etal}{et~al.\ }
\newcommand{\magsec}{mag~arsec$^{-2}$}

\newcommand{\eg}{e.g.,\ }

\def\muv{$\mu_{\mbox{v}}$}

\documentclass[cup5b]{caps}
\usepackage{graphicx}
\usepackage{amssymb}
\usepackage{ociwsymp3e}  

\HeadText{J. J. Feldmeier et al.}  

\begin{document}

\pagenumbering{arabic}


%
\author[]{J. J. FELDMEIER$^{1}$, J. C. MIHOS$^{1}$, H. L. MORRISON$^{1}$, 
P. HARDING$^{1}$, and N. KAIB$^{1}$
\\(1) Case Western Reserve University, Cleveland, OH, USA}
\chapter{Using Intracluster Light to Study Cluster Evolution}
\begin{abstract}
We present some early results from our deep imaging survey of galaxy clusters
intended to detect and study intracluster light (ICL).  From our
observations to date, we find that ICL is common in galaxy clusters,
and that substructure in the ICL also appears to be common as well.  
We also discuss some initial comparisons of our imaging results to 
high-resolution numerical simulations of galaxy clusters, and give
avenues for future research.
\end{abstract}

\section{Introduction}

The concept of intracluster starlight (ICL), or stars between the galaxies
in galaxy clusters is not a new one: it was first proposed over
50 years ago (Zwicky 1951).  However, progress in studying ICL
has been slow due to its low surface brightness, which is less 
than 1\% of the brightness of the night sky 
(see V{\'i}lchez-G{\'o}mez 1999, Feldmeier 2000 for reviews).
This is unfortunate, because ICL is a powerful probe of the evolution 
of galaxies in clusters (Dressler 1984), and of cluster evolution overall.

In the last six years however, the study of ICL has increased dramatically.
Numerous individual intracluster stars have been detected in nearby and
distant galaxy clusters (Arnaboldi \etal 1996; Theuns \& Warren 1997;
Feldmeier, Ciardullo \& Jacoby 1998; Ferguson, Tanvir \& von Hippel 1998; 
Durrell \etal 2002; Feldmeier \etal 2003; Arnaboldi \etal 2003; 
Gal-Yam \etal 2003).  At the same time, deep imaging of clusters with
CCD detectors have also detected the intracluster light 
(Uson, Boughn, \& Kuhn 1991; V\'ichez-Gomez, Pell\'o, \& Sanahuja, 1994; 
Bernstein \etal 1995; Gonzalez \etal 2000).
Detections of tidal debris arcs (Trentham \& Mobasher 1998; 
Gregg \& West 1998; Calca\'neo-Rold\'an \etal 2000) 
in clusters have shown that at there is significant substructure
in the ICL, and that the production of ICL is ongoing.

Encouraged by these results, we have undertaken a deep imaging
survey of galaxy clusters, intended to quantify the properties of ICL
as a function of environment, and overall galaxy cluster properties.
From our deep imaging, with careful attention to systematic errors 
(\eg Morrison, Boroson, \& Harding 1994; Morrison \etal 1997), we are able 
to measure the ICL to faint surface brightnesses many magnitudes 
below that of the night sky ($\mu_{\mbox{v,ICL}}$ $\approx$ 26--28).  
In tandem with the observations, we are constructing numerical 
simulations of galaxy clusters in a cosmological context, similar
to those of Dubinski (1998).  

Here, we present the results of our deep imaging survey to date.  
We note that there are several other searches for ICL underway that have 
complementary goals (see Gonzalez \etal this volume; Krick \etal this volume).

\section{The Survey}

For our initial survey, we focus on observing Abell clusters 
(Abell, Corwin, \& Olowin 1989) at large distance classes 
(5--6, corresponding to z $\approx$ 0.1--0.2).  This is
to ensure that the entire cluster is contained in our field-of-view 
for good sky subtraction, but the clusters are not so 
distant that cosmological $(1+z)^4$ surface brightness dimming is 
a major effect.  We are also observing a few nearby clusters from the MKW/AWM 
(Morgan, Kayser, White 1975, Albert, White, \& Morgan 1977) as a comparison 
sample.  Our goal is to gather a {\it representative} sample of 
clusters with different richness, Bautz-Morgan and Rood-Sastry types.  
We have observed ten clusters so far (ACO 84, 98, 545, 801, 1234, 1413, 
1553, 1914, 2443 and MKW 7), and plan to observe 4--5 more before 
completing this initial survey.  

We observe using the KPNO 2.1m, and image through the Washington 
{\it M} filter, which is similar to {\it V} but contains fewer night sky 
emission lines.  We use the ultra-deep surface photometry techniques of 
Morrison, Boroson, \& Harding (1994) for our observations and
data reduction.  We spend half of our telescope time constructing 
our dark sky flats (which are flat to 0.3\% on all scales 
in the worst case), and the other half observing the clusters.  
We then carefully mask out all objects in our frames, using a 
combination of the DAOPHOT (Stetson 1987), SExtractor 
(Bertin \& Arnouts 1996) and our own software.  After detecting the 
ICL, we construct an error model for each cluster that includes 
all sources of error, both random and systematic.  Our data has a 
signal-to-noise of five (including systematic errors; see Feldmeier 
\etal 2002 for an example of the error model) at a surface 
brightness of 26.5 mag~arcsec$^{-2}$.  The signal-to-noise ratio 
approaches unity at \muv = 28.3 \magsec.  

The results of the first two clusters have 
been published in Feldmeier \etal (2002): work is ongoing 
on the other clusters.  A mosaic of the first five clusters is 
shown in Figure~\ref{fig:mosaic}.  Once this initial survey is complete,
we plan to observe nearby clusters using the CWRU Burrell Schmidt, which
is currently being optimized for ultra-deep surface photometry.  

\section{Results}

From the clusters we have observed so far, it seems that diffuse 
intracluster light and intracluster tidal debris is common in 
galaxy clusters.  
Fig~\ref{fig:mkw7} shows an example of a tidal plume superimposed over 
the brightest cluster galaxy in the cluster MKW~7.  Similar tidal features 
can be seen in Abell~1914 (see the boxed regions in Figure~\ref{fig:mosaic}),
and in several other clusters.  Although our sample is still 
small, it seems that the clusters that are cD-dominated (Abell~1413, MKW~7) 
have smoother, more regular ICL than those in clusters that 
do not have a cD galaxy (ACO 1234, 1553 \& 1914).  In these non cD 
clusters, the ICL follows an irregular distribution, and is not well 
correlated with galaxy density.

Tidal features of the kind we have detected can be seen in high-resolution 
cluster N-body simulations (\eg Moore \etal 1996; Dubinski 1998;
Dubinski, Murali, \& Ouyed 2001; see also Willman, this volume), 
but thus far we have seen few long tidal debris arcs.  
Plume-like intracluster debris structures seem to be more common in 
the clusters we have surveyed thus far.  However, it is clear from 
the simulations that the vast majority of tidal debris seen in cluster 
simulations has a surface brightness much lower than our limit of \muv = 26.5 
\magsec~ for the initial observations.  The structures that we 
have observed are likely to be the brightest features in each cluster. 

Another interesting facet of our observations is the nature 
of cD galaxy envelopes, which are characterized by an excess of diffuse light 
(compared to an  R$^{1/4}$ law) at large radius.  The origin of cD 
envelopes remains unclear: are they formed in the 
initial stages of cluster collapse (Merritt 1983, 1984), or 
later, as galaxies continue to fall in the cluster and become 
tidally stripped?  The accepted view of cD envelopes (Schombert 1992) is that 
cDs in rich galaxy clusters have large extended cD envelopes, while brightest 
cluster galaxies in poorer clusters do not.  In our observations of
Abell~1413 and MKW~7, we have found the opposite behavior: MKW~7, a poor
cluster has a strong cD envelope, while Abell 1413 has a weaker 
cD envelope than previously measured.  Since other researchers have
also found cDs in rich clusters with pure R$^{1/4}$ laws 
(\eg Gonzalez 2000), this may signify that the earlier photographic 
surface photometry of cDs may need to be re-examined.

\section{Future Work}

Our initial survey will be completed by the end of 2003: we then
plan to begin observations on the Burrell Schmidt.  The wide
field of view of the Schmidt ($\approx$ 1.5 degrees) will allow
us to observe more nearby galaxy clusters, where much more is known
about their properties.  We also plan to run additional 
large-scale galaxy cluster simulations over a range of cosmologies 
and initial conditions so that we can make detailed comparisons between 
our imaging survey and theoretical results.  

\begin{figure}
    \centering
    \includegraphics[width=6cm,angle=0]{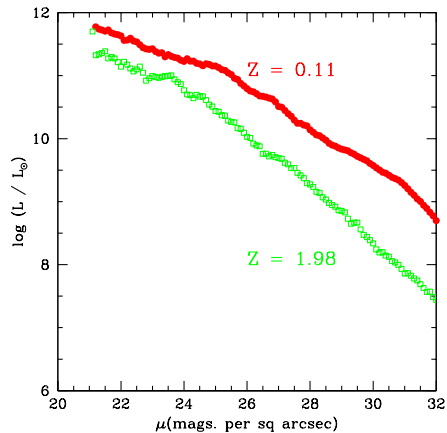}
    \caption{A surface brightness histogram of all starlight from a
N-body simulation by Dubinski (1998).  As can be clearly seen, as the cluster
evolves, more starlight is accreted (causing an offset between the two
histograms), and starlight is stripped from the high-surface brightness 
galaxies, and re-distributed into lower surface brightness ICL features.  
This stripping causes the surface brightness histogram slope to 
become more shallow over time.  This relation, which is in principle 
observable, may allow us to place limits on the dynamical age of 
galaxy clusters.}
    \label{fig:lmu}
  \end{figure}

A final goal is to create quantitative metrics for the ICL that can be 
used on simulations and observations.  In Figure~\ref{fig:lmu} we show
one potential metric: the amount of stellar luminosity as a function 
of surface brightness, applied to the cluster simulation of Dubinski (1998)
at two different redshifts.  In this relation, dynamically evolved clusters 
have a shallower slope to their ``surface brightness distribution function''
than dynamically younger clusters.  From comparison of multiple 
simulations, the intrinsic scatter in this slope at fixed redshift 
is low ($\approx$ 4\%).  More testing is needed, but the 
results seem promising.

  \begin{figure}
    \centering
    \caption{Images of five clusters observed in this survey.
From left to right and top to bottom, they are: Abell~1413, MKW~7,
Abell~1914, Abell~1234, and Abell~1553.  The last three clusters 
show only the central section of the image;
these clusters show a wealth of ICL substructure.  Many of the
brighter galaxies at the center of each cluster lie within a low
surface brightness common envelope, and there are clearly defined
tidal features, most noticeably in Abell~1914 (denoted by the boxes).  
This in contrast to the first two clusters, which had less 
substructure (Feldmeier \etal 2002). }
    \label{fig:mosaic}
  \end{figure}

  \begin{figure}
    \centering
    \caption{Our residual image for MKW~7, after the best-fitting
elliptical model of the cD + ICL has been subtracted.  The 
black ellipse shows where the measured surface brightness has
a signal-to-noise greater than five.  A large tidal plume is apparent
leading from the center of the image to the right (south), 
and up (west) of the galaxy's nucleus.  The magnitude of this plume
is approximately equal to a small galaxy ($M_{\mbox{v}} \sim -17$)}
\label{fig:mkw7}
  \end{figure}

\begin{thereferences}{}

\bibitem{aco} 
Abell, G. O., Corwin, H. G., \& Olowin, R. P. 1989, \apjs, 70, 1

\bibitem{awm} Albert,
C. E., White, R. A., \& Morgan, W. W.\ 1977, \apj, 211, 309 

\bibitem{1996arna} Arnaboldi, M., \etal 1996, \apj, 472, 145

\bibitem{}Arnaboldi, M.~et al.\ 2003, \aj, 125, 514 

\bibitem{bern1995} Bernstein, G. M., Nichol, 
R. C., Tyson, J. A., Ulmer, M. P., \& Wittman, D. 1995, \aj, 110, 1507

\bibitem{sex1996} Bertin, E.\ \& 
Arnouts, S.\ 1996, \aas, 117, 393

\bibitem{cr2000}
Calc\'aneo-Rold\'an, C. , Moore, B. , Bland-Hawthorn, J. , 
Malin, D.,  \& Sadler, E. M. 2000, \mnras, 314, 324

\bibitem{dress1984}Dressler, A. 1984, \annrev, 22, 185

\bibitem{dub1998}Dubinski, J. 1998, \apj, 502, 141

\bibitem{dub2001} Dubinski, J., Murali, C.,
\& Ouyed, R. 2001, unpublished preprint

\bibitem{durr2002} Durrell, P., Ciardullo, R., 
Feldmeier, J. J., Jacoby, G. H., \& Sigurdsson, S. 2002, \apj, 570, 119

\bibitem{thesis}Feldmeier, J. J. 2000, Ph.D. Thesis, Penn State University 

\bibitem{ipn1} 
Feldmeier, J. J., Ciardullo, R., \& Jacoby, G. H. 1998, \apj, 503, 109 

\bibitem{ipn2}Feldmeier, J.J., Ciardullo, R., Jacoby, G.H., Durrell, P.R.
2003, \apjs, 145, 65

\bibitem{icl1} Feldmeier, J.~J., 
Mihos, J.~C., Morrison, H.~L., Rodney, S.~A., \& Harding, P.\ 2002, 
\apj, 575, 779 

\bibitem{ftv1998} 
Ferguson, H. C., Tanvir, N. R., \& von Hippel, T. 1998, \nat, 391, 461 

\bibitem{gal1} Gal-Yam, A., Maoz, D., Guhathakurta, P., \& Filippenko, A. 2003,
\aj, 125, 1087

\bibitem{gon2000} Gonzalez, A. H., Zabludoff, A. I., 
Zaritsky, D., \& Dalcanton, J. J. 2000, \apj, 536, 561 

\bibitem{gregg1998} 
Gregg, M. D., \& West, M. J. 1998, \nat, 396, 549

\bibitem{mer1983} Merritt, D. 1983, \apj, 264, 24

\bibitem{mer1984} Merritt, D. 1984, \apj, 276, 26

\bibitem{harass} 
Moore, B., Katz, N., Lake, G., Dressler, A., \& Oemler, A. 
1996, \nat, 379, 613

\bibitem{mkw} Morgan, W. W., Kayser, S., \& White, R. A. 1975, \apj, 199, 545

\bibitem{sb1994} Morrison, H. L., 
Boroson, T. A. \& Harding P., 1994, \aj, 108, 1191

\bibitem{sb1997} Morrison, H. L., Miller, E. D., 
Harding, P., Stinebring, D. R., \& Boroson, T. A. 1997, \aj, 113, 2061 

\bibitem{schombert1992} Schombert, J.\ 1992, ASSL 
Vol.~178: Morphological and Physical Classification of Galaxies, 53

\bibitem{1987PASP...99..191S} Stetson, P.~B.\ 1987, \pasp, 99, 191

\bibitem{t1997} Theuns, T., \& Warren, S. J. 1997, \mnras, 284, L11

\bibitem{tren1}Trentham, N. \& Mobasher, B. 1998, \mnras, 293, 53

\bibitem{uson1991b} 
Uson, J. M., Boughn, S. P., \& Kuhn, J. R. 1991, \apj, 369, 46

\bibitem{vg1994} V\'ichez-Gomez, R., 
Pell\'o, R. \& Sanahuja, B. 1994, \aa, 283, 37

\bibitem{vg1999}V{\'i}lchez-G{\'o}mez, R. 1999, ASP Conf. Ser. 170: The 
Low Surface Brightness Universe, 349

\bibitem{z1951} Zwicky, F. 1951, \pasp, 63, 61

\end{thereferences}
\end{document}